\documentclass{aa} 
                                   
\usepackage{graphicx}
\usepackage{amsmath}
\usepackage{color,soul, colortbl} 
\usepackage[varg]{txfonts}

\definecolor{Gray}{gray}{0.9}
\graphicspath{{pictures/}}

\begin{document}
\title{Migration of massive planets in accreting disks}

\author{C. D\"urmann \and W. Kley}

\institute{Institute of Astronomy and Astrophysics, Universit\"at T\"ubingen, Auf der Morgenstelle 10, 72076 T\"ubingen, Germany \\
          \email{christoph.duermann@uni-tuebingen.de}
         }

\date{Received ; accepted }
 
\abstract
{}
{Massive planets that open a gap in the accretion disk are believed to migrate with exactly the viscous speed of the disk, a regime termed type II
migration. Population synthesis models indicate that standard type II migration is too rapid to be in agreement with the observations.
We study the migration of massive planets between $2\times10^{-4}$ and $2\times10^{-3}~M_\odot$,
corresponding to 0.2 to 2 Jupiter masses $M_\mathrm{J}$
to estimate the migration rate in comparison to type II migration.}
{We follow the evolution of planets embedded in two-dimensional, locally isothermal disks with non-zero mass accretion, which
is explicitly modelled using suitable in- and outflow boundary conditions to ensure a specific accretion rate. 
After a certain relaxation time we release the planet and measure its migration through the disk and the dependence on
parameters, such as viscosity, accretion rate, and planet mass. We study accreting and non-accretion planets.}
{The inferred migration rate of the planet is determined entirely by the disk torques acting on it and is completely independent of the viscous
inflow velocity, so there is no classical type II migration regime. 
Depending on the local disk mass, the migration rate can be faster or slower than type II migration.
From the torques and the accretion rate profile in the disk we see that the gap formed by the planet does not separate the inner from the outer disk as necessary for type II migration, rather gas crosses the gap or is accreted onto the planet.}
{}


\maketitle

\section{Introduction}
The interaction of the growing protoplanet with the ambient disk leads to a change in the orbital elements of the planet.
The most important for the overall evolution of the planet is the change in semi-major axis, i.e. the migration of the planet.
The topic of planet-disk interaction and the orbital evolution of planets has been covered in a few recent reviews
\citep{2012ARA&A..50..211K, baruteau2013recent, 2013arXiv1312.4293B}, and here we present only a brief summary of the relevant issues.
Depending on the mass of the planet, different regimes of migration are distinguished. Most important are the two limiting regimes
of type I and type II migration.

Type I migration occurs for low mass planets (with a mass less than about 50~$M_\oplus$) that do not open a gap in the disk,
and can be treated in the linear regime. Here the total torque is given by the effects of the spiral density waves
(Lindblad torques) and the corotation torques generated by the gas flow in the coorbital horseshoe region. 
In contrast to the Lindblad torque that quite generally leads to inward migration
\citep{1997Icar..126..261W}, the corotation torque suffers from saturation effects making it strongly dependent on
on the magnitude of viscosity and thermal diffusion. These can significantly alter the migration speed and may even reverse
the direction of the migration. Analytical formulae to account for these effects have been presented 
by \citet{paardekooper2010torque} for adiabatic disks and \citet{2011MNRAS.410..293P} for disks with thermal diffusion. 

Type II migration occurs for massive planets with masses comparable to $M_\mathrm{J}$ or larger. Because of angular momentum deposition in the
disk, those planets open an annular gap in the protoplanetary disk at the location of the planet where the density is significantly reduced.
In this case the migration speed is slowed down from the linear rate because of the reduced mass available near the planet.
During the migration process the gap created by the planet will have to move with the planet through the disk.
Hence, it is often assumed that in an equilibrium situation the gap moves exactly with the viscous accretion velocity of the disk,
and that the planet is locked in the middle of the gap to maintain torque equilibrium 
\citep{ward1982tidal,lin1986tidal,1997Icar..126..261W}. 

Direct numerical simulations of planet-disk interaction typically place the planet in a non-accreting disk and keep the planet at a fixed
position to analyse the torque density distribution from which the migration speed can then be calculated, see \citet{2012ARA&A..50..211K} for
references. Calculations, where gap-opening Jupiter-type planets were allowed to move through the disk as given by the torques
acting on them, have been performed for isothermal disks by \citet{2000MNRAS.318...18N} for two-dimensional (2D) simulations.
They studied non-accreting and accreting planets where mass from inside the planet's Roche lobe was added to the planet mass.
Both cases lead to similar results indicating that a massive planet could migrate within $10^5$ years from about 5 au all the way
to the center. 
In full 3D radiative simulations even faster migration rates were found, though only for planets with a maximum mass 
of about 0.6$M_\mathrm{J}$ \citep{2010A&A...523A..30B}.
Numerical calculations of migrating massive planets were carried out by \citet{edgar2007giant,edgar2008type}. 
Using a constant kinematic viscosity he came to the result that there is no constant migration rate as predicted for the type II regime.
In a related study \citet{2014MNRAS.444.2031P} analyzed the influence of planetary motion on the type I migration regime.
He found that in non-isothermal disks the regime of outward migration can be larger than that estimated for fixed planets.

The migration of massive planets in evolving disks has been analysed by \citet{2007MNRAS.377.1324C} for 2D isothermal disks.
In their models, the planets were placed in a disk with an initial Gaussian profile that evolves under a constant kinematic
viscosity. Their results show that only planets that carve a deep gap in the disk experience genuine type II migration and
move with the viscous speed of the disk. This is the case either for very massive planets or low viscosity disks,
with a Reynolds number larger than about $5 \times 10^5$.  In the case of only partial cleared gaps,
i.e. for less massive planets and larger viscosity, they found a reduced migration speed and in some cases even outwards migration,
an effect due to the very effective action of the corotation torques. In addition, because of the positive density slope
at the location of the planet in their simulations, the Lindblad torques are strongly reduced, which strengthens the effect. 

Following this global study there have been recent models of planets embedded in disks with a given (constant) mass accretion rate.
\citet{bitsch2014stellar}  focussed on the regime of earth-mass planets in an irradiated accretion disk with a given $\dot{M}$ rate
to determine the regimes of inward and outward migration. They modelled axisymmetric disks with vertical structure and used the
formula by \citet{2011MNRAS.410..293P} to determine the migration properties of the embedded planets.
\citet{fung2014empty} investigated the structure of gaps created by stationary planet in an accreting disk, but they did not include planet migration.
In their 2D isothermal studies they showed that despite the presence of a massive, gap opening planet, which was not allowed to accrete any material,
the disks reached an equilibrium state with a constant mass accretion rate. Obviously, in this configuration the mass flow across the gap 
equals exactly the imposed disk accretion rate, $\dot{m}$.
The timescale of type II migration was analysed with respect to the agreement with the orbital properties of the observed massive extrasolar planets
by \citet{2013ApJ...774..146H}. They argued that standard type II migration is too rapid and mechanisms need to be found to slow it down.

Recently, \citet{duffell2014type} looked at the migration of planets in accretion disks using an alternative, different approach. 
Instead of moving the planet according to the disk torques, they pulled a Jupiter mass planet through a zero $\dot{m}$ non-accreting disk,
measured the torques acting on the planet, and compared the resulting migration rate with the actual pull rate of the planet.
In this way they obtained possible equilibrium solutions of the migration speed of a planet through a disk. In particular, they found
that the possible equilibrium migration speed of the planet is independent of the viscous speed and can be lower and smaller than type II migration,
depending on the local surface density of the disk.

Hence, it appears that the issue of type II migration is presently not resolved. Theoretically, it is clear that a planet can only be moved by the disk
torques acting on it. On the other hand, disk material moves under the action of viscous torques and its local inflow speed is directly proportional to the
viscous torque. If a planet migrated exactly in the type II regime, then the disk torque and viscous torque should be equal. 
In this paper, we make a new attempt at the problem of type II migration and study the migration of planets in 2D isothermal disks
with a constant mass accretion rate. In a first step, we construct constant $\dot{m}$ disks with stationary planets of different masses and
then, in a second step, we move the planets according to the torques acting on it, 
measure their migration rate, and compare it to the type II migration rate.

Our setup is described in Sect.~\ref{sec:setup}, and in Sect.~\ref{sec:fixed_planet} we present our results for fixed planets in accreting disks.
In Sect.~\ref{sec:migrating_planets} we move the planets according to the torques acting on them, and compare this in detail to type II migration
in Sect.~\ref{sec:typeII}. The results are discussed in the final section \ref{sec:conclusion}.
 
\section{Setup}
\label{sec:setup}
To study the migration of planets in disks, we assume that the disk is geometrically thin and simplify to a two-dimensional (2D) approximation.
The disk is assumed to be locally isothermal and driven by an $\alpha-$type viscosity. In addition we model explicit mass accretion through the disk.
For our calculations we use the \texttt{NIRVANA}-code \citep{ziegler1997nested, ziegler1998nirvana} in a 2D setup. The star is located at the centre of a
cylindrical coordinate system, which covers a radial range of  1.56 to 15.6 au corresponding to \mbox{$r=0.3\ldots 3.0$} in code units where the unit of length
is given by $r_0=5.2$\,au. In the following we will specify distances in code units unless specified otherwise. In the azimuthal direction we cover a complete annulus from $0$ to $2\pi$. 
In our standard resolution the domain is covered by \mbox{$251\times 583$} cells.
The planet is placed on a circular orbit at a distance of 5.2 au corresponding to $r=1.0$.

To speed up the initial relaxation we, first calculated the equilibrium of the disk, without the planet, with a reduced resolution of \mbox{$(N_r,N_\phi)=(101,233)$}. After $10^6$ time steps, corresponding to about 5000 orbits, these results were interpolated to our standard-resolution grid. Then the model could adapt to this new resolution for additional $10^5$ time steps (about 290 orbits) before the planet was released. The planet then could move freely in the disk, and change its semi-major axis and eccentricity.
In addition, we performed another set of similar calculations with a lower resolution of \mbox{$(N_r,N_\phi)=(135,405)$} and extending from $r=0.2$ to $2.0$. Both calculations give similar results, after 1000 orbits the positions of the planets differ by around 5\,\%.

For our studies we varied the viscosity by changing the $\alpha$-parameter, 
the accretion rate (and thus the disk surface density, see Eq.~\eqref{eq:sigma} below), and the planetary mass.
The used parameter space with the highlighted standard model, is shown in table \ref{tab:parameter_space}.
We limit our planet mass to a maximum of 2 $M_{\text{Jup}}$ (or $q =0.002$) because for larger $m_{\text{p}}$ the outer disk becomes eccentric
\citep{2006A&A...447..369K} and migration properties change \citep{2006ApJ...652.1698D}. 

\begin{table}
\centering
\renewcommand{\arraystretch}{1.2}
\begin{tabular}{l l l}
\hline\hline
\multicolumn{1}{c}{$\alpha$} & \multicolumn{1}{c}{$\dot{m}$} & \multicolumn{1}{c}{$q$} \\ \hline
 & $2\times10^{-9}$ &  \\
 & $5\times10^{-9}$ &  \\
 & $1\times10^{-8}$ &  \\
 & $2\times10^{-8}$ & 0.0002 \\
0.001 & $5\times10^{-8}$ & 0.0005 \\
\rowcolor[gray]{.9}0.003 & $1\times10^{-7}$ & 0.001 \\
0.01 & $2\times10^{-7}$ & 0.002 \\

\hline
\end{tabular}
\caption{Parameter space used in our calculations. The highlighted values define our standard model. In the standard-resolution (\mbox{$251\times 583$}) models only one parameter was varied while the other two were constant. In the low-resolution models (\mbox{$135\times 405$}) the whole parameter space with $\dot{m}>10^{-8}$ was covered.
The values of $\alpha$ and $q=M_\mathrm{p}/M_\odot$ are dimensionless and $\dot{m}$ is given in $M_\odot /\mathrm{yr}$ in the whole article.
}
\label{tab:parameter_space}
\end{table}

\subsection{Initial and boundary conditions}
The goal of this work was to set up a disk with a steady accretion flow through the disk. Hence, matter has to be fed into the domain at the other boundary of the disk (at $r_{max}$), transported to the inner disk, leaving the domain at $r_{min}$. 
The local accretion rate through a disk is given by
\begin{equation}
\dot{m}=-2\pi r \Sigma u_r \,. 
\end{equation}
It depends on the surface density, $\Sigma$, and the radial velocity $u_r$, of the gas. 
In equilibrium for a constant $\dot{m}$ the viscous accretion velocity is given by
\begin{equation}
u_r^\mathrm{visc}=-\frac{3}{2}\frac{\nu}{r} \,,
\label{eq:u_r}
\end{equation}
where $\nu$ is the kinematic viscosity and $r$ the radial distance to the central star. 
This result can be obtained for stationary accretion disks with constant accretion rate, derivations can be found in textbooks
such as \citet{1992apa..book.....F} or \citet{2010apf..book.....A}. 
In our case of an isothermal disk, we use the $\alpha$-viscosity prescription, $\nu = \alpha c_s H$, \citep{shakura1973black} with the disk scale height $H=h r$, where the relative disk thickness is constant, $h=0.05$.
With the isothermal sound speed, $c_s = H\Omega_\mathrm{K}$, this can be written as
\begin{equation}
u_r^\mathrm{visc}=-\frac{3}{2}\frac{\alpha c_s H}{r} = -\frac{3}{2}\alpha h^2 r \Omega_\mathrm{K},
\label{eq:u_r-alpha}
\end{equation}
where $\Omega_K=\sqrt{GM_*/r^3}$ is the Keplerian orbital frequency with the stellar mass $M_*$ and the gravitational constant $G$.
With given $\dot{m}$ and $\alpha$ this leads in equilibrium to the surface density 
\begin{equation}
  \Sigma = \frac{1}{3\pi}\frac{\dot{m}}{\alpha h^2 r^2 \Omega_\mathrm{K}} = \frac{\dot{m}}{3\pi \alpha h^2 \sqrt{G M}} \, r^{-1/2} = \Sigma_0 r^{-1/2}.
\label{eq:sigma}
\end{equation} 
From radial equilibrium (pressure gradient balanced by centrifugal force and gravity) the angular velocity is given by
\begin{equation}
  u_\phi = \sqrt{1-\frac{h^2}{4}} \, r \Omega_\mathrm{K}.
\label{eq:u_phi}
\end{equation}
Thus, for our initial conditions we use $u_r^\mathrm{visc}$ from Eq.~\eqref{eq:u_r-alpha}, $\Sigma$ from Eq.~\eqref{eq:sigma} and
$u_\phi$ from Eq.~\eqref{eq:u_phi}, with prescribed values of $\dot{m}$, $\alpha$ and $h$.

At the outer boundary gas enters the computational domain with $u_r^\mathrm{visc}$, and $\Sigma_0$ is fixed according to the chosen $\dot{m}$. 
For the standard model with $\alpha=0.003$ and $\dot{m}=10^{-7} M_\odot/\mathrm{yr}$, the initial disk density at $r=1$ is $\Sigma_0=878 \mathrm{g\,cm^{-3}}$.
At the inner boundary only the radial velocity is fixed 
according to Eq.~\eqref{eq:u_r-alpha}, material then flows out with the local density at $r_{min}$. The azimuthal velocity $u_\phi$ is held fixed, according to Eq.~\eqref{eq:u_phi}, at both the inner and the outer boundary. In the azimuthal direction, we enforce periodic boundaries.
For the gravitational potential, we use $\epsilon$-smoothing with $\epsilon=0.6H$ \citep{2012A&A...541A.123M}. To prevent density fluctuations due to wave reflection at the radial boundaries the radial and azimuthal velocity is damped towards its azimuthal mean value near the boundaries on a timescale of $\tau_\mathrm{damp} \approx \Omega_\mathrm{K}^{-1}$ with $\Omega_\mathrm{K}$ taken at the inner and outer boundaries.

\subsection{Accretion onto the planet}
To study the influence of mass accretion by the planet on the migration speed, we performed additional
simulations in the two extreme cases of non-accreting planets and maximally accreting planets. 
Both cases are of course unrealistic, but reality lies somewhere in between.
To model maximal accretion, we implemented the accretion scheme devised by \citet{kley1999mass} where at each time step the gas density inside the Roche Lobe \mbox{$R_\mathrm{R} \approx a (q/3)^{1/3}$} of the planet is reduced by a factor of $1-f_\mathrm{red}\Delta t$ with $f_\mathrm{red}=1/2$. 
In our simulations, the mass removed in this way is not added to the planet mass.

\section{The disk structure for fixed planets}
\label{sec:fixed_planet}
\begin{figure}[t]
\includegraphics[width=9cm]{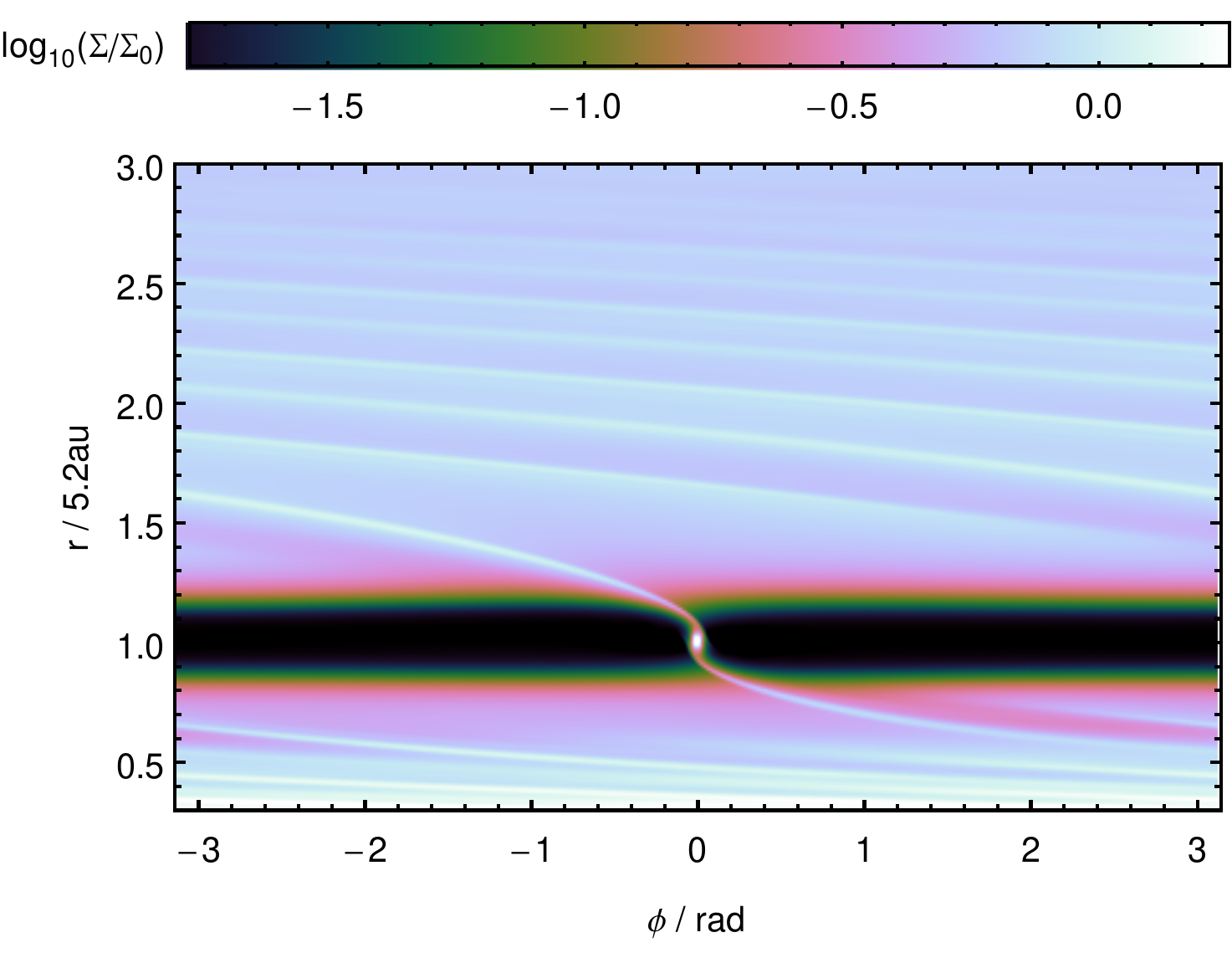}
\caption{Surface density of a fixed-orbit, non-accreting planet calculation for the standard model with $(\alpha,M_\mathrm{p}/M_*,\dot{m})=(0.003,0.001,10^{-7})$ after 2500 orbits. The planet is in the middle of a stable gap and the accumulation of mass near the planet is clearly visible. The spiral arms are damped at the inner and outer boundary so no reflections are seen.}
\label{fig:sigma2d}
\end{figure}
Before starting the calculations with moving planets, we constructed equilibrium cases where the planets are not allowed to move and remain at their initial
locations. This allows us to estimate the size and depth of a gap as a function of planet mass and the viscosity.
The surface density of this calculation for the standard model after 2500 orbits is shown in Fig.~\ref{fig:sigma2d}. There is a very steady gap edge and stable spiral arms resulting from the perturbation of the planet.

\begin{figure}[t]
\includegraphics{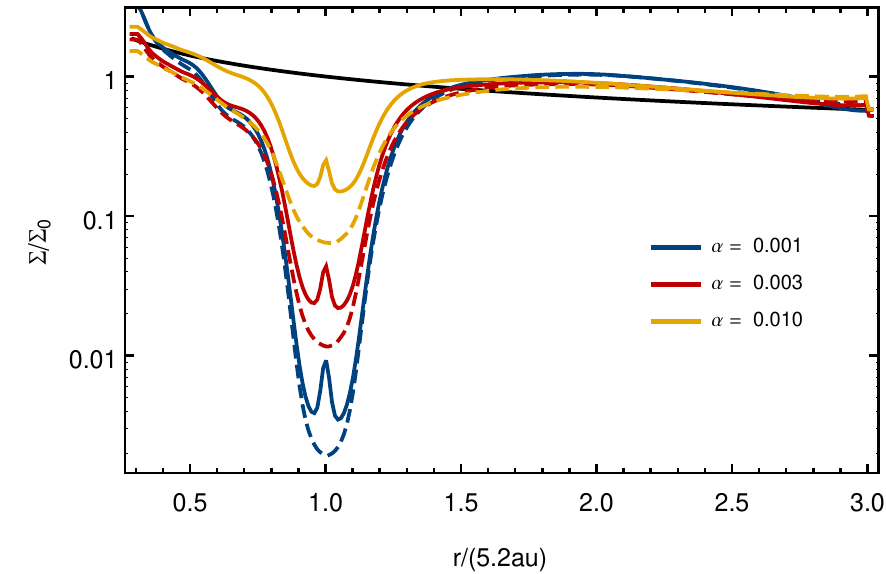}
\caption{Azimuthally averaged surface density for a non-accreting (solid line) and an accreting (dashed line) planet with $q=0.001$, for
different values of the viscosity parameter $\alpha$. The black line indicates the initial density profile at the beginning of the simulation.
The red line corresponds to the standard model.}
\label{fig:gap-alpha}
\end{figure}
\begin{figure}[t]
\includegraphics{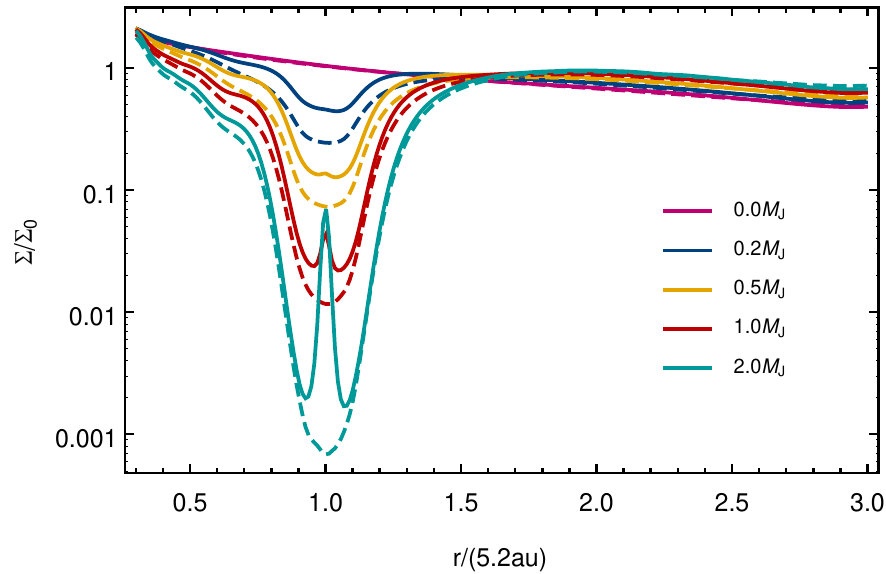}
\caption{Azimuthally averaged surface density for non-accreting (solid line) and accreting (dashed line) planets and $\alpha=0.003$, for different
values of the planet mass. The red line corresponds to the standard model.}
\label{fig:gap-planetmass}
\end{figure}
The azimuthally averaged surface density in the case of stationary planets is shown for different values of $\alpha$ (in Fig.~\ref{fig:gap-alpha}) and $q$ (in Fig.~\ref{fig:gap-planetmass}). As expected, the gaps are deeper for lower viscosity and higher planet mass. While a more massive planet causes a much wider gap, decreasing the viscosity primarily deepens the gap. In isothermal disks, for stationary planets, the surface density distribution is identical for different values of the accretion rate, and hence $\dot{m}$ has no influence on the gap profiles that are therefore identical.
The gap depth depends on the accretion onto the planet. For non-accreting planets the gap has a clearly visible density bump at the position of the planet. If accretion onto the planet is allowed, the bump disappears and the gap becomes deeper because gas is accreted onto the planet and removed from the simulations. 

\begin{figure}[t]
\includegraphics{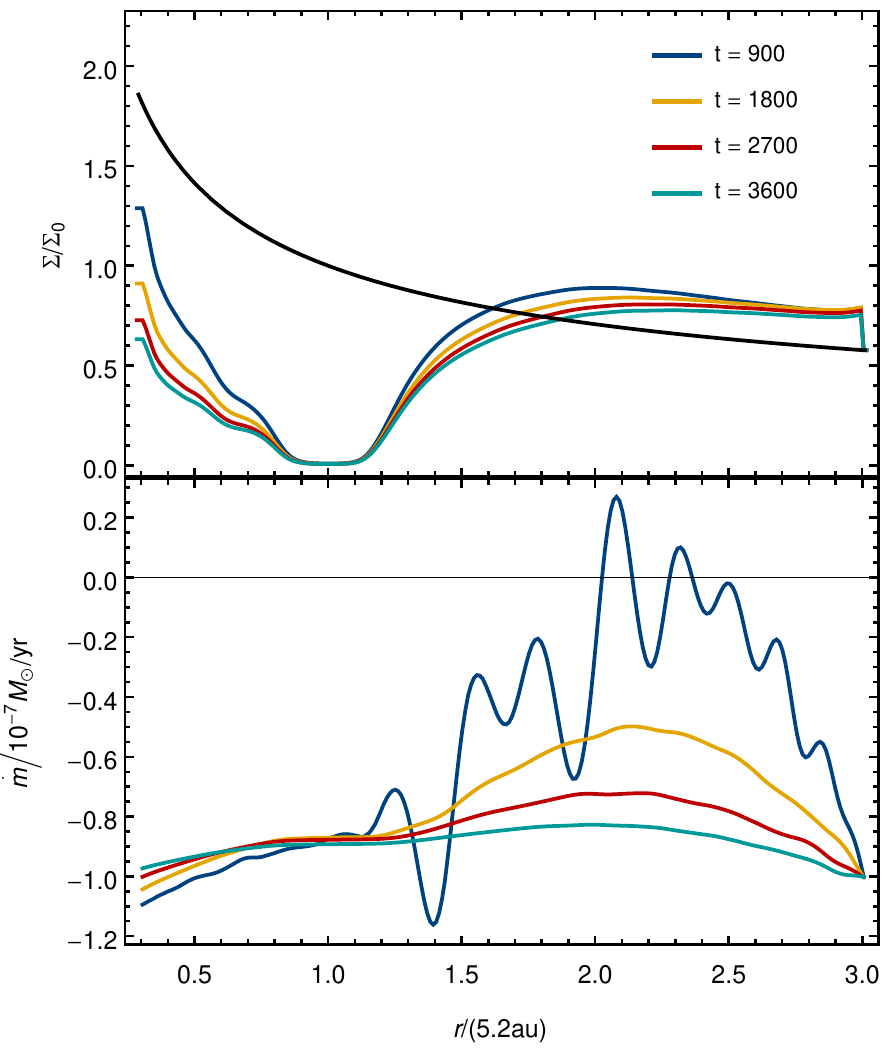}
\caption{Azimuthally averaged surface density profile (top) and the migration rate profile (bottom) in case of $(\alpha,q,\dot{m})=(0.003,0.001,10^{-7})$ as in Fig.~\ref{fig:sigma2d}. The planet is at $r=1.0$. The profiles are taken at different times after the beginning of the calculation. The black line indicates the initial density profile at the beginning of the simulation before the planet was added.}
\label{fig:gap-accretion-profiles}
\end{figure}

In the outer disk region, for $r>1.7$, the surface density is increased over the case without a planet and the amount increases with planet mass. This is because the gas initially in the gap is pushed to the inner and outer regions of the disk, because of angular momentum transfer to the disk.
Because, in the outer disk, the gas cannot leave the computational domain the surface density must be increased, the effect being stronger for more massive planets with a deeper gap.  Hence, there occurs a jump in the surface density at the outer boundary.
This is also clearly seen in Fig.~\ref{fig:gap-accretion-profiles} where no logarithmic scale is used.
This jump in $\Sigma$ at $r_{max}$ has no influence on our results on the migration properties and it disappears if larger $r_{max}$ are used.

The evolution towards the equilibrium state is shown in Fig.~\ref{fig:gap-accretion-profiles} where the radial dependence of the surface density and local accretion rate, $\dot{m}(r)$, are displayed at different times. The migration rate is obtained by integrating the mass fluxes in the advection routine of the code 
over the $\phi$-direction yielding a radial profile of accretion. This ensures consistency with the numerical calculations.
In the bottom panel, it is clearly visible that the equilibrium is not yet reached after 3600 orbits. This is because the accretion rate is extremely sensitive to all kinds of perturbations. The equilibrium viscous inflow speed is given by $u_r/c_s \approx \alpha h$, which is about $10^{-4}$ for our standard model. Hence even small pressure
perturbations lead to large variations in the radial velocity, compared to the viscous speed. 
Thus, although the accretion rate profile is not flat, the changes in the surface density after 2700 and 3600 orbits are very small, and the gap is already fully developed. Obviously, in a steady-state solution the gas must be able to cross the gap. The inner and outer disk are connected and there is no significant pile up at the outer gap edge or depletion at the inner gap edge. 

\begin{figure}[t]
\includegraphics{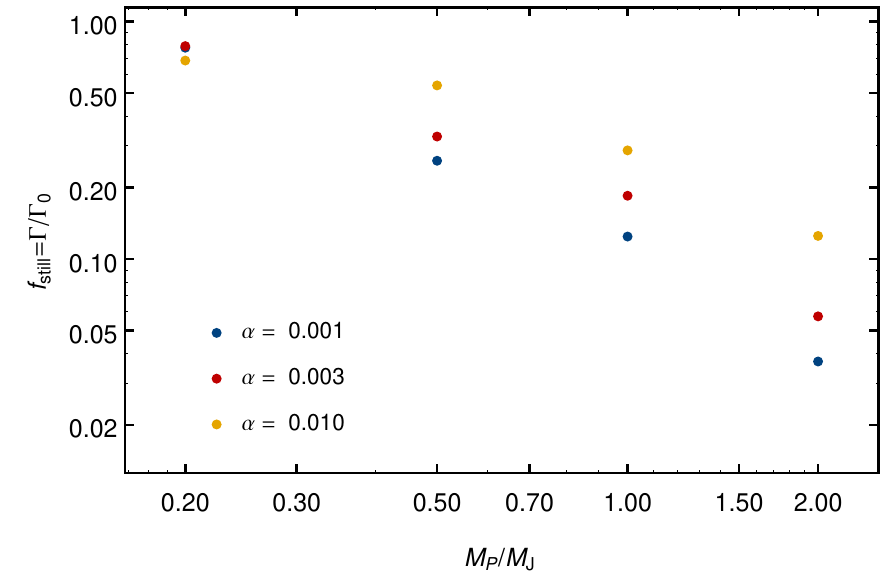}
\caption{Normalized torques acting on planet with fixed orbits at the end of the relaxation as a function of the planet mass. 
The data is taken from the low-resolution calculations. For isothermal disks the normalized torques are independent of the disk density. }
\label{fig:torque-mass-fixed}
\end{figure}
\begin{figure*}
\centering
\includegraphics[width=\textwidth]{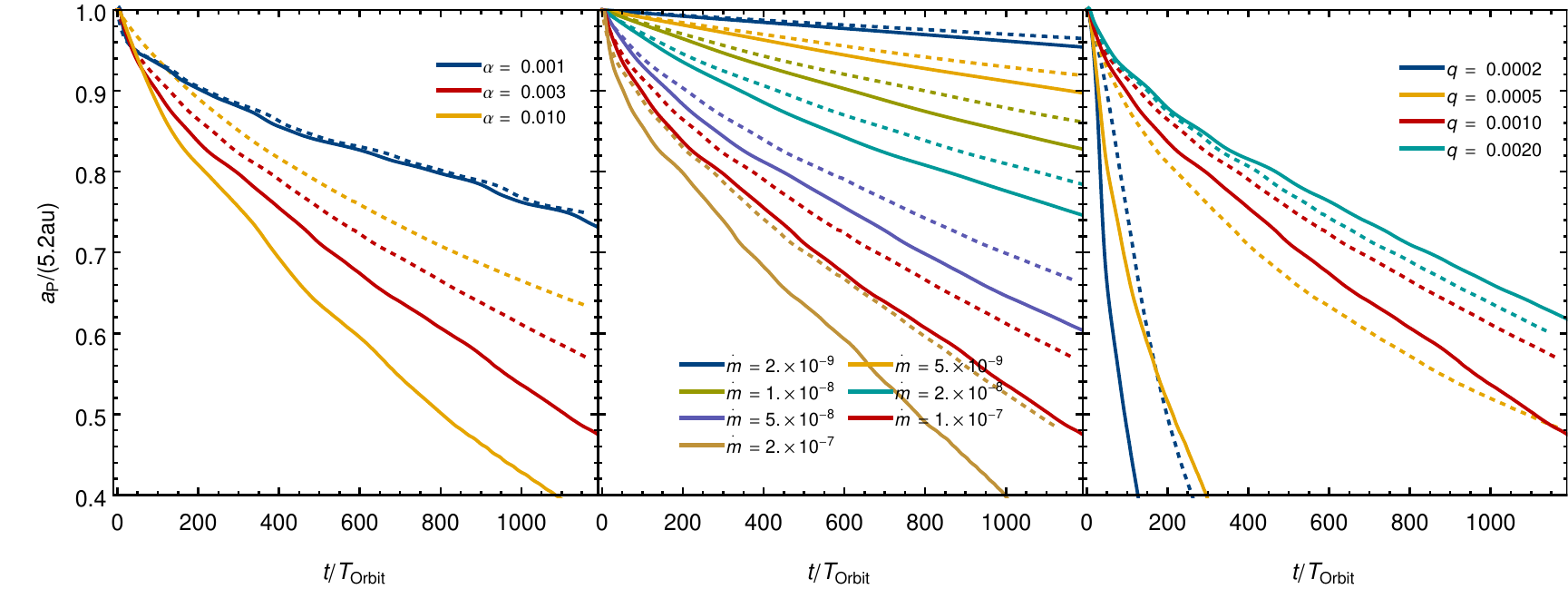}
\caption{Planet migration tracks for different model parameter.
Starting from the standard model (in red) with $(\alpha,\dot{m},q) = (0.003,10^{-7},0.001)$, individual parameters have been varied while keeping two others fixed. We varied $\alpha$ (left panel), $\dot{m}$ (middle panel), and $q$ (right panel). 
The full and dotted lines correspond to non-accreting and accreting planets, respectively.}
\label{fig:migration-tracks}
\end{figure*}

\subsection{The torque acting on the planet}
To estimate the expected migration rate of planets in disks with mass flow, we calculate the torques generated by the density perturbations
in the disk as induced by the planet.
In Fig.~\ref{fig:torque-mass-fixed} we display the obtained torques as a function of planet mass $M_\mathrm{P}$ for different value of the viscosity.
Here, we used the standard torque normalization \citep[see e.g.][]{paardekooper2010torque}
\begin{equation}
\label{eq:Gamma0}
\Gamma_0 =-\Sigma_\mathrm{p} \Omega_\mathrm{K}^2 a_\mathrm{p}^4 \, \left(\frac{q}{h}\right)^2,
\end{equation}
where $\Sigma_\mathrm{p}$ is the unperturbed surface density at the location of the planet and $a_\mathrm{p}$ its semi-major axis.
This definition of $\Gamma_0$ takes the linear results into account for low mass planets that are subject to type I migration with $\Gamma \propto M_\mathrm{p}^2$, Massive planets open gaps in the disk that reduce the torque, and a different scaling is expected.

To strengthen the point that these torques are measured for fixed planets that are not allowed to migrate, we introduce the correction
factor $f_{\text still}$ and write for the torque acting on the planet
\begin{equation}
            \Gamma = f_\mathrm{still} \, \Gamma_0 \,.
\end{equation}

As seen from Fig.~\ref{fig:torque-mass-fixed}, the total torque is indeed reduced for larger planet masses.
For small planet masses, a flattening is expected as one approaches the type I regime. In our case, the planet masses are still
too high, but this effect becomes marginally visible for the higher viscosity models, as expected.
On the other hand, there is no visible dependence on variations in the disk mass or the accretion rate (which are coupled here), 
because in isothermal disks the disk mass is just a scaling factor that depends linearly on the specified $\dot{m}$ value.

\section{Migrating planets}
\label{sec:migrating_planets}
After having analysed the equilibrium state for fixed planets, we release now the planets and evolve their orbital elements according to the action of the disk.
We measured the migration tracks for many different parameters (see table \ref{tab:parameter_space}) for more than 1000 orbits for accreting and non-accreting planets. The migration tracks are shown in Fig.~\ref{fig:migration-tracks} for a variety of model parameters.
All planets migrate inwards at a rate that decreases with time. 
At the beginning of the migration, shortly after the release of the planet, there is a short phase of rapid inward migration that then slows down. 
This transient phenomenon is caused by the unavoidable small mismatch of a disk structure with a stationary planet in comparison to the evolving
case.

In cases where $M_\mathrm{P} \lesssim M_\mathrm{D} = \Sigma_\mathrm{P} a_\mathrm{P}^2$, with $\Sigma_\mathrm{P}$ being the undisturbed surface density at the position of the planet $a_\mathrm{P}$, there is no slowing down, rather we observe very fast type III migration, such as for $q \leq 0.0005$ (yellow and blue curve) in the right panel of Fig.~\ref{fig:migration-tracks}. Type III migration sets in later for accreting planets.

\subsection{Mass accretion onto the planet}
Our implementation of the mass accretion onto the planet is very simple, and, therefore, we only used this method to obtain an idea what changes are expected in case of an accreting planet. Because the algorithm takes away the gas near the planet at a high rate, it serves as an upper limit of the possible accretion onto the planet. 
We found that the general behaviour of accreting planets is similar to the non-accreting cases. As seen in Fig.~\ref{fig:migration-tracks} accreting planets migrate more slowly in most cases. The reason is the reduced density in the vicinity of the planet, which leads to weaker torques. Although the contribution of the planet envelope inside $0.8$ Hill-Radii is not considered when calculating the torque
\citep[see][]{2008A&A...483..325C}, the region affected by the accretion is much bigger and thus important for the torques. This effect of accretion is similar to what has been seen in \citet{2000MNRAS.318...18N} who added the accreted material to the dynamical mass of the planet, however. This will lead to an even slower migration of the planet due to the increased inertia.

\subsection{Gap profiles during the evolution}
\begin{figure}[t]
\includegraphics[width=9cm]{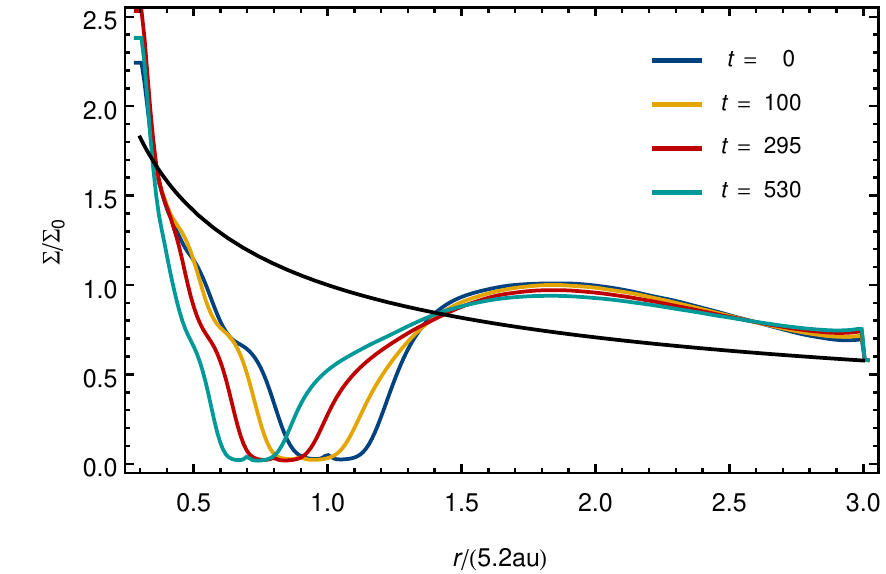}
\caption{Global gap profiles at four different times during migration of the planet for the standard model with $q=0.001$ and $\dot{m}=10^{-7}$.
The black line indicates the unperturbed density profile without the planet. 
}
\label{fig:gap-profiles-global}
\end{figure}

The gap profile has an important impact on the migration of a planet. Because the disk is depleted near the planet's orbit, these regions cannot contribute to the torques on the planet. 
In Fig.~\ref{fig:gap-profiles-global}, the change in global disk density is displayed for four different snapshots during the evolution of the planet
in our standard model.
At the inner edge of the gap, we observe a density pile up that initially increases with the planet moving inwards closer to the inner edge. But when the planet reaches $r \approx 0.7$ at $t=530$, it decreases again because the gap approaches $r_{min}$, and at the inner boundary an outflow condition with a given radial velocity is enforced.
At the outer gap edge we see the gas lagging behind the movement of the gap. The gap moves $\Delta r=0.3$ in 530 orbits. The viscous timescale $\tau_\mathrm{visc}=\Delta r^2/\nu$ for the gas to cover this distance is about 2070 orbits. It is therefore obvious that the planet and the gap move faster than the viscous speed of the gas, which demonstrates that the gap must be dynamically created during the planet's inward motion.
The inward motion of the planet and the gap can only be maintained because gas can cross the gap faster than the viscous timescale.

\begin{figure}[t]
\includegraphics{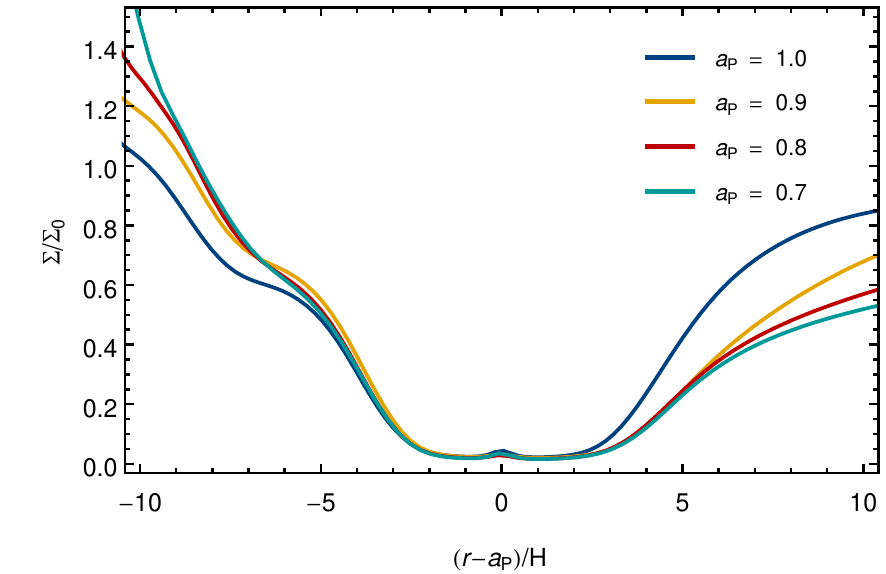}
\caption{Gap profiles for the 
standard model taken at different positions of the planet in the disk during its migration path. 
The gap is rescaled using the local disk scale height, $H$, at the position of the planet. 
The curves correspond directly to those shown in Fig.~\ref{fig:gap-profiles-global}.
}
\label{fig:gap-profiles-position}
\end{figure}

To obtain a more detailed view on the local disk structure in the gap region, we display
in Fig.~\ref{fig:gap-profiles-position} the gap at different times during the evolution of the planet in the disk. 
The gap does not change significantly during the migration through the disk if the width is rescaled with the local disk scale height $H(a_\mathrm{p})$
at the actual position of the planet. Farther away from the planet the surface density varies because of the different positions in the disk.

\begin{figure}[t]
\includegraphics[width=9cm]{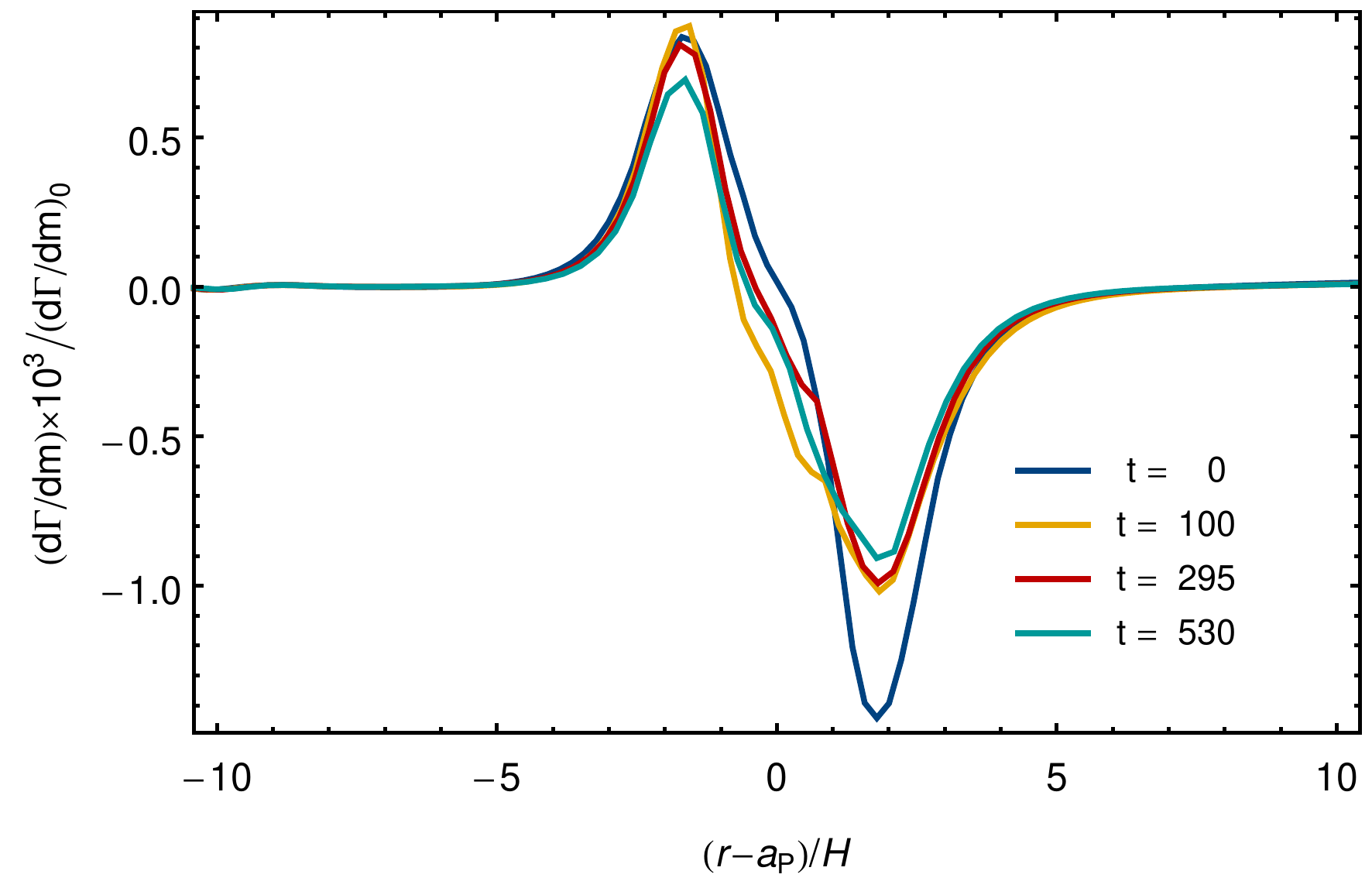}
\caption{Specific torques for the same model as in Fig.~\ref{fig:gap-profiles-position} at the same times and respective positions.}
\label{fig:torques-gap}
\end{figure}

In Fig.~\ref{fig:torques-gap} the specific torques are displayed where we use the normalization of \citet{d2010three},
\begin{equation}
\left(\frac{\mathrm{d}\Gamma}{\mathrm{d}m}\right)_0=q^2h^{-4} a_\mathrm{P}^2 \Omega_\mathrm{P}^2 \,.
\end{equation}
The biggest contribution is generated within a $5H$ wide region inside and outside the planet. Apart from the initial adjustment of the gap to the moving planet (see Fig.~\ref{fig:gap-profiles-position}) the specific torques reach an equilibrium profile after 100 orbits. As expected from very similar local gap profiles (in Fig.~\ref{fig:gap-profiles-position}), the specific torques do not depend strongly on the position of the planet in the disk. This will be of importance later in explaining the overall migration properties of the planets
(see Sect. \ref{sec:typeII}).

\begin{figure}[t]
\includegraphics{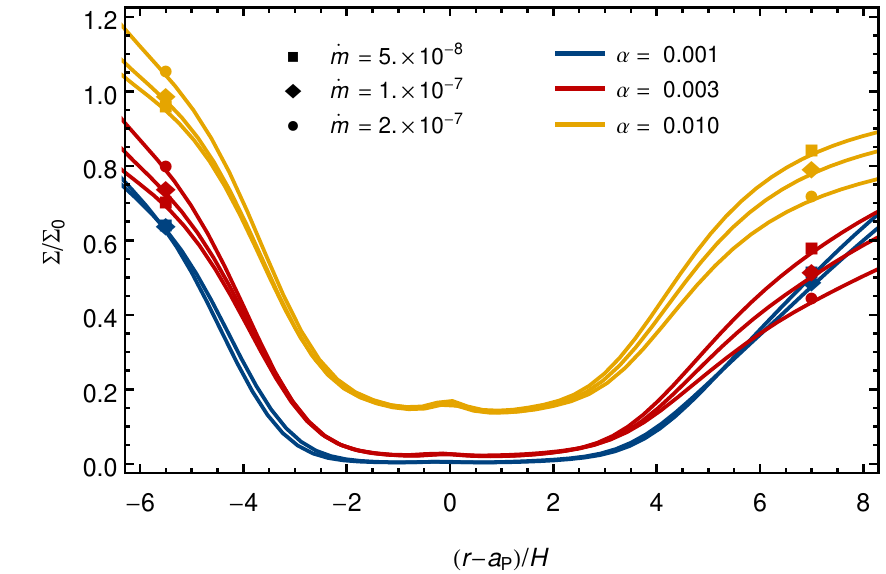}
\caption{Gaps of a $q=0.001$ planet with different accretion rates 100 orbits after the release of its orbit. The gap is rescaled according to the disk scale height at the position of the planet, which varies for different values of $\alpha$ and $\dot{m}$. The gap profiles are taken from the low-resolution calculations.}
\label{fig:gap-profiles-alpha-accretion}
\end{figure}

As shown in Fig.~\ref{fig:gap-profiles-alpha-accretion} the gaps, of course, depend on the viscosity. For a fixed $\alpha$-value, the gap structure near the planet does not change very much, even for different disk masses. 
Effects farther away from the planet $\gtrsim 4 H$ are results from the different positions in the disk as well as the movement of the gap through the disk with different velocities (see Fig.~\ref{fig:migration-tracks}). 
At the inner gap edge the surface density is increased for the higher accretion rates and at the outer gap edge for the smaller accretion rate. This is because the inward migration rate of the planet is faster for a denser disk (and thus a higher accretion rate, see Fig.~\ref{fig:migration-tracks}) and the gap edges cannot follow as fast as the planet moves. A higher migration rate therefore leads to a stronger gap deformation. 
The gap depth depends on $\alpha$ in the same manner as for the static planets in Fig.~\ref{fig:gap-alpha}.

\subsection{Flow across the gap}
\begin{figure}[t]
\includegraphics{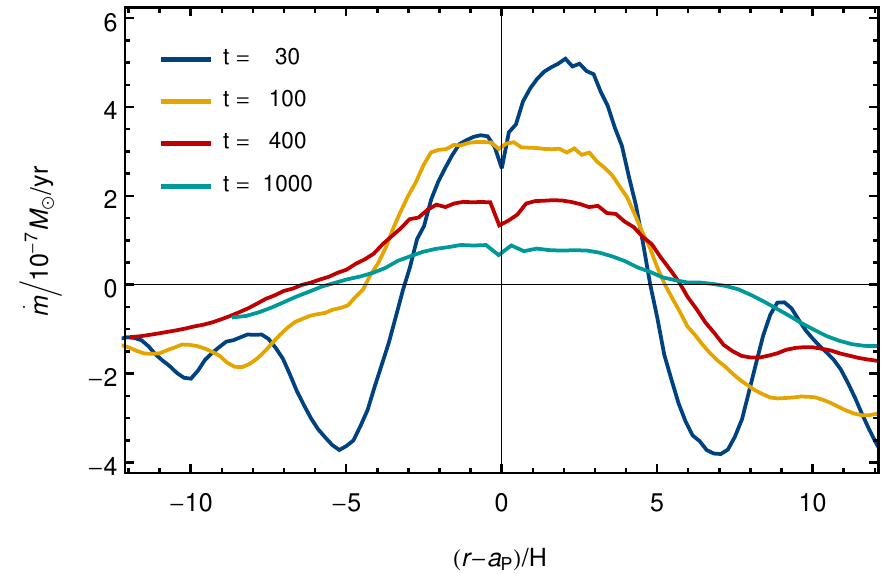}
\caption{Azimuthally averaged local accretion rate, $\dot{m}(r)$, at different times during the inward migration of the planet for the standard model ($q=0.001$, $\dot{m}=10^{-7}$, $\alpha=0.003$). The plot is similar to Fig.~\ref{fig:gap-accretion-profiles}, which has the same parameters, but now for moving planets. To compare the accretion rate near the planet the position is rescaled to units of the disk scale height at the position of the planet. The labels refer to the elapsed time (in initial orbits) after the release of the planet.}
\label{fig:accretion-profiles-freeOrbit}
\end{figure}

Above we have argued that the fast migration speed of the planet implies that mass has to cross the gap. Now we analyse this process
in more detail. In Fig.~\ref{fig:accretion-profiles-freeOrbit} we display the local mass accretion rate through the disk, $\dot{m}(r)$, 
similar to Fig.~\ref{fig:gap-accretion-profiles}, but now for a migrating planet at different times after the release.
At early times, when the planet is moving inwards very fast, the gas is flowing outwards instead of inwards. Even at later times, when the migration rate has slowed down, at the position of the planet the direction of the gas flow is still reversed. 
This effect is caused by the motion of the planet through the disk, which disturbs the local disk structure and leads locally, at
the location of the planet, to a transfer of mass from the inner to the outer disk.
The mass transfer is positive because the planet moves faster than the viscous accretion velocity $u_r^\textrm{visc}$. 
As shown above, the gap is bound to move with the planet, but the planet does not  drive the disk structure as whole. In consequence, gas from the inner disk, not able to move inwards faster than $u_r^\mathrm{visc}$, has to move outwards, which means crossing the gap in outward direction. 
For a planet moving more slowly than the viscous speed (for low disk densities) the radial flow remains always negative.

If the disk were separated and gas could not cross the gap, the inner disk would pile up and the region adjacent to the planet in the outer disk would be depleted because the gas could not follow the planet's movement. The more massive inner disk would then yield higher positive torques and slow down or even reverse migration. Thus, seeing transport of the gas outwards is a clear sign that the planet is not bound to the viscous accretion velocity and moves independently.
\section{Comparing to type II migration}
\label{sec:typeII}
\begin{figure}[t]
\includegraphics{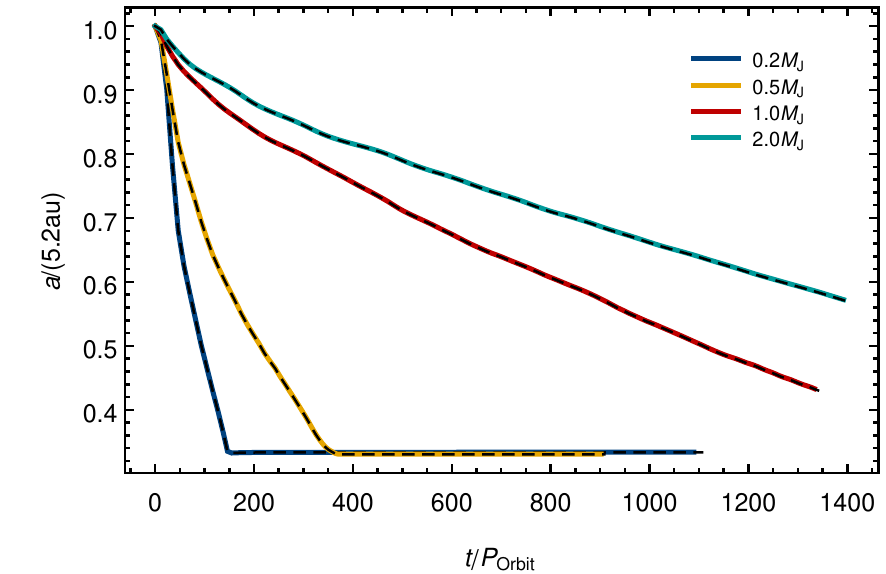}
\caption{ Migration tracks of planets embedded in disks for different planet masses, for given viscosity and disk accretion rate
$(\alpha,\dot{m})=(0.003,10^{-7})$. 
Coloured lines: Theoretical tracks calculated by integrating the measured torques.
The dashed black lines are the corresponding observed migration tracks in the simulations. }
\label{fig:integrated-migration}
\end{figure}
\begin{figure}[t]
\includegraphics{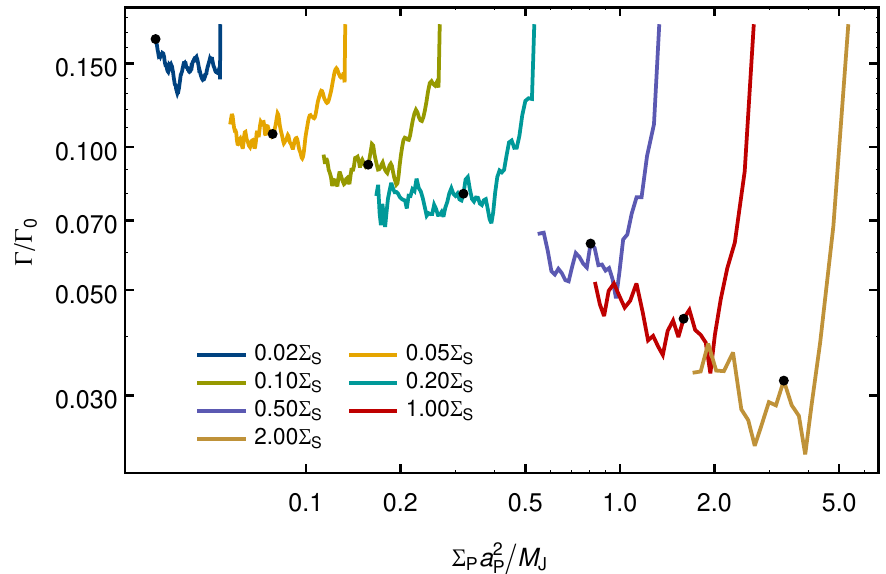}
\caption{Normalized torques acting on a Jupiter mass planet as a function of the local disk mass, with $\alpha=0.003$ and $q=0.001$. 
Shown is the evolution of the torques during the migration process, each track starts at the rightmost point with the highest
value of $\Sigma_\mathrm{P} a_{\mathrm{P}}^2$.
The black dots mark the point when the planet has reached $a_\mathrm{P}=0.7$. During the inward migration the local disk mass $\Sigma_{\mathrm{P}}a_{\mathrm{P}}^2$, where $\Sigma_\mathrm{P}$ is the surface density in the undisturbed surface density profile at planet position $a_\mathrm{p}$, decreases. 
The models differ in their initial disk density $\Sigma_0$ varied by the disk accretion rate $\dot{m}$, where $\Sigma_\mathrm{S}$ corresponds to $\dot{m}=10^{-7}$.
The torques are averaged over 50 orbits.
}
\label{fig:torques-diskmass}
\end{figure}
After having obtained the migration track of planets in disks with net mass flux we are now in a position to
compare these to the type II migration speed of a viscously driven disk. 

In any case, the migration of a planet can only be driven
by the gravitational torques exerted on it by the disk material. To prove this statement, we plot in
Fig.~\ref{fig:integrated-migration} the evolution of the planet's semi-major as expected from the calculated total torque, $\Gamma$,
during its motion through the disk (coloured lines).
The migration track $a(t)$ is obtained by integrating the formula
\begin{equation}
\label{eq:adot}
\dot{a} = \frac{2\Gamma}{M_\mathrm{p}  a_\mathrm{P} \Omega_\mathrm{K}}.
\end{equation}
As seen in Fig.~\ref{fig:integrated-migration}, the observed migration (black dashed lines) agrees perfectly with the
calculated migration tracks (coloured lines)  throughout the whole simulation. 
Hence, the migration is indeed completely determined by the torques from the disk, as physically expected.

Having shown that the torques determine the migration of the planet, we plot in
Fig.~\ref{fig:torques-diskmass} the normalized torques using the same normalization, $\Gamma_0$ as before
in Fig.~\ref{fig:torque-mass-fixed}. 
Note that now the normalization factor $\Gamma_0$ is a function of the planet's position, see Eq.~\eqref{eq:Gamma0}. 
In Fig.~\ref{fig:torques-diskmass} the torques are displayed during the evolution of the planet. 
The initial location
refers to the rightmost points in the individual curves. They all start at the same height $\Gamma/\Gamma_0 = 0.18$ 
as given by the fixed planet resultis, see Fig.~\ref{fig:torque-mass-fixed}. During the inward migration the value of $\Sigma_\mathrm{p} a_\mathrm{p}^2$
decreases. The black dot in each curve corresponds to the time where the planet has reached $a_\mathrm{p} = 0.7$. 

Let us first compare the findings to the linear case, for small mass planets that do not open gaps. 2D simulations yield the following relation for the
total Lindblad torque on planets \citep[see][]{2002ApJ...565.1257T,paardekooper2010torque}
\begin{equation}
\frac{\Gamma_\mathrm{L}}{\Gamma_0} = 3.2 + 1.468 \, p \,,
\label{eq:torque-paardekooper}
\end{equation}
where $p$ is the exponent of the surface density profile, with $\Sigma(r) \propto r^{-p}$, and so $p=0.5$ in our case.
This yields $\Gamma_\mathrm{L}/\Gamma_0 = 3.934$. 
Expression \eqref{eq:torque-paardekooper} was derived for much less massive planets in the regime of type I migration and is not directly applicable to our calculations. The idea is to compare the migration speed of massive planets to the linear type I regime.
As shown in Fig.~\ref{fig:torques-diskmass} the torques are about 20 to 100 times smaller than the linear results, 
a consequence of the gap. Additionally, the results show a clear
dependence on the density of the disk. In the linear case one would expect a constant ratio  $\Gamma/\Gamma_0$ for all disk densities.

This fact is not apparent overall, but after the disk and the gap have adapted to the moving planet, the torques become constant for each individual curve, i.e. for each disk density. 
This implies that the normalized torques for a given planetary mass and $\alpha$-value are completely defined by the disk density alone. Therefore, we introduce a slow down factor for migrating planets $f_\mathrm{mig}$, which gives the factor by which the torques acting on the moving planet are smaller than those for the same planet on a fixed orbit. Because for gap-opening planets the fixed-orbit torques are again smaller by a factor $f_\mathrm{still}$ than the type I torques $\Gamma_0$ (see Fig.~\ref{fig:torque-mass-fixed}) we obtain the following expression for the torque of a moving planet
\begin{equation}
\Gamma = f_\mathrm{mig}  \, f_\mathrm{still}  \, \Gamma_0 \,.
\label{eq:fmig-fstill}
\end{equation}

\begin{figure}[t]
\includegraphics{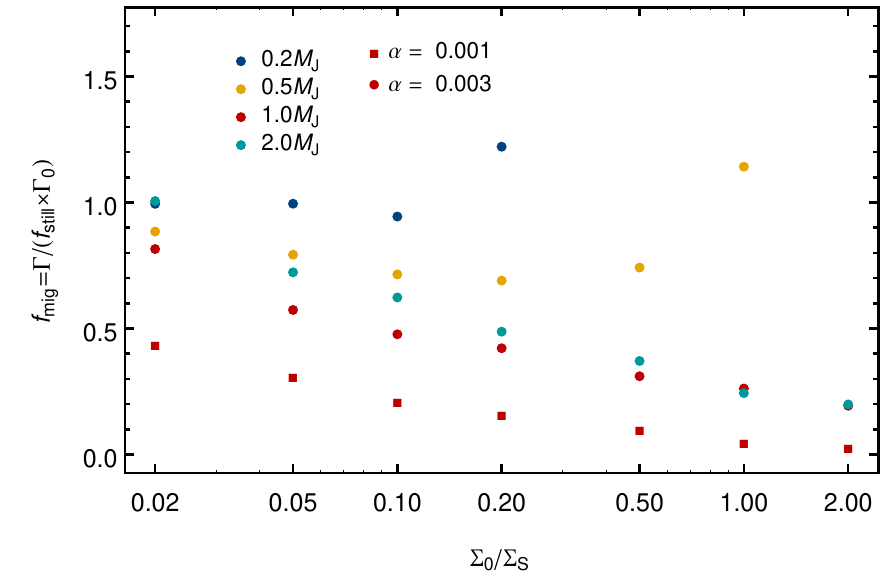}
\caption{Slow down factor $f_\mathrm{mig}$ depending on the disk density for different planetary masses and viscosities.
Here, the surface density is given in units of $\Sigma_\mathrm{S}$, as defined in Fig.~\ref{fig:torques-diskmass}. 
The circles are results for $\alpha = 0.003$ and the squares for $\alpha = 0.001$.
}
\label{fig:fmig}
\end{figure}

In Fig.~\ref{fig:fmig} we show the slow down factor $f_\mathrm{mig}$ for various parameters. For small disk densities the slow down factor decreases with increasing disk density. For light planets with $M_\mathrm{P}<M_\mathrm{J}$ this changes for $\Sigma_0/\Sigma_\mathrm{S}$ between $0.1$ and $0.5$ and the slow down factor strongly increases to values much bigger than one, which marks the onset of type III migration.
For the highest disk densities, the slow down factor for the lighter planets are in fact out of the plot range and reach values up to $f_\mathrm{mig}=12$.
In case of smaller viscosity $f_\mathrm{mig}$ decreases, but for the same planetary mass the overall structure seems to remain the same.

\begin{figure}[t]
\includegraphics{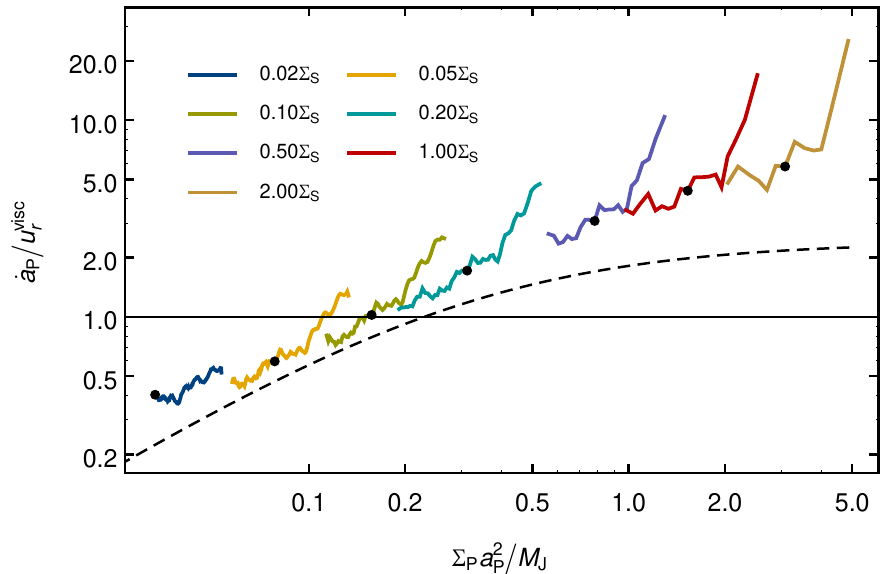}
\caption{Migration rate of a Jupiter mass planet normalized by the viscous accretion velocity. Apart from the different $y$-axis the plot is similar to Fig.~\ref{fig:torques-diskmass}.
The horizontal line corresponds to $u_r^\mathrm{visc} = 1$. The dashed line shows a fit of \citet{duffell2014type}, but they used constant viscosity equal to $\alpha=0.01$ at $r=5.2$\,au, so their fit is normalized with $u_r^\mathrm{visc}$ for that $\alpha$.
}
\label{fig:dota-mdot}
\end{figure}
In Fig.~\ref{fig:dota-mdot} we compare the obtained migration speed to the type II migration rate as given by the
viscous inflow speed $u_r^\mathrm{visc}$. 
In the case of classical type II migration, the curves should be independent from disk mass and identical to the viscous inflow speed,
apart from the adjustment in the beginning.
However, the results show a clear dependence with disk mass.
Obviously, the actual migration of a planet is not coupled to the viscous evolution in the disk, as also pointed out recently
by \citet{duffell2014type}. 
Qualitatively, the relationship between disk mass and migration rate is similar to that found by \citet{duffell2014type} found by different means
(see dashed curve in Fig.~\ref{fig:dota-mdot}). In our case the inward migration is faster than theirs.
For light disks with $\Sigma_\mathrm{D}=\Sigma_\mathrm{P}a_\mathrm{P}^2 < 0.2 M_\mathrm{J}$ only,
we find that migration becomes smaller than the viscous speed. 
Here, it is important to note that the viscous speed for our constant $\dot{m}$ models scales directly as $u_r^\mathrm{visc} \propto a_p \Omega$
the quantity that \citet{duffell2014type} used in their plots.

\begin{figure}[t]
\includegraphics{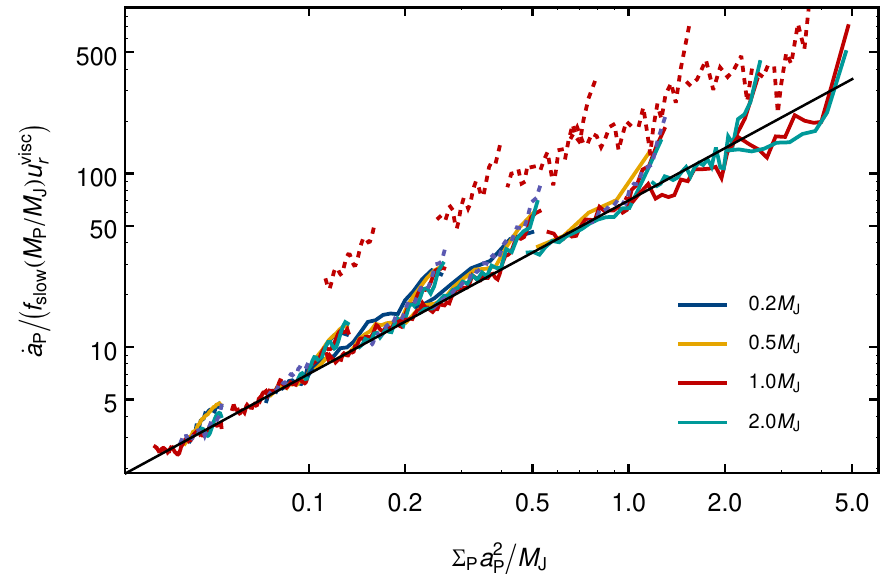}
\caption{Normalized migration rates of planets with different masses between $0.2$ and $2~M_\mathrm{J}$. 
The migration rate is scaled with the slow down factor and planet mass (see relation~\ref{eq:fslow}). The slope of the black line is 1. The dotted line corresponds to $\alpha=0.001$. Accreting planets are indistinguishable from the non-accreting planets and therefore not shown here. Otherwise the plot is similar to Fig.~\ref{fig:torques-diskmass}.
}
\label{fig:migrate-planetmass}
\end{figure}

From our results we obtained for moving planets that the normalized torque $\Gamma/\Gamma_0$ was constant during the evolution of a
planet (see Fig.~\ref{fig:torques-diskmass}). This implies that the torque reduction factors $f_\mathrm{mig}$ and $f_\mathrm{still}$ do not depend
on the local disk mass, and are only functions of the planet mass and the disk viscosity.
Using Eq.~\eqref{eq:adot} for the migration rate and the scaling for $\Gamma_0$, we obtain for each planet track 
\begin{equation}
    \dot{a} \propto  a_\mathrm{P}^{3/2} \, f_\mathrm{slow} \, \Sigma_\mathrm{P} M_\mathrm{P} \,,
\end{equation}
where $f_\mathrm{slow} = f_\mathrm{mig} \, f_\mathrm{still}$.
Using the relation $u_r^\mathrm{visc} \propto r^{-1/2}$, as inferred from Eq.~\eqref{eq:u_r-alpha} for constant $H/r$, one finds
\begin{equation}
\label{eq:fslow}
    \frac{\dot{a}}{f_\mathrm{slow} u_r^\mathrm{visc} M_\mathrm{P}}  \propto   a_\mathrm{P}^{2} \Sigma_\mathrm{P} \,.
\end{equation}
This relation is plotted in Fig.~\ref{fig:migrate-planetmass} and indeed the data lie on a straight line, for all the 
models with different $\dot{m}$. The data for an accreting planet fall onto the same line. In case of different viscosity there
 is a shift due to the change in $f_\mathrm{slow}$, but it follows the same trend.
 
\section{Summary and conclusions}
\label{sec:conclusion}
We have studied the migration of massive planets in locally isothermal disks with net mass flow, $\dot{m}$, through them. 
In our initial step, keeping the planet at a fixed location, we showed that despite the presence of the planet, the disk can nevertheless transport the 
full $\dot{m}$ through the disk and the gap. We analysed the structure of the gap for different planet masses and viscosities. 
As expected, the depth of the gap increases with planet mass and is reduced with increasing viscosity.
Our gap structure and depths for these models are in very good agreement with the empiric formulae presented in \citet{fung2014empty}.
Additionally, the gap opening criterion by \citet{crida2006width} gives a suitable condition if there is a considerable gap depth,
$\approx 0.1$ of the ambient density.
For large planet mass the normalized torque on the planet 
strongly decreases (see Fig.~\ref{fig:torque-mass-fixed}),
in contrast to the linear rate where $\Gamma/\Gamma_0$ has no dependence on planet mass. This is a consequence of the increased gap
width with a reduced mass that can drive the planet.

Upon releasing the planet we find rapid inward migration, often faster than the typical viscous inward drift of the disk material.
As expected, the planet moves exactly according to the disk torques acting on it. A migration independent of the disk's drift speed
implies that material can cross the gap region during the migration process. 
We performed simulations without and with mass accretion onto the planet where, in the latter case, the accretion rate was near the maximum
possible rate \citep{kley1999mass}. As expected, because of the reduced mass in the vicinity of the planet, the migration rate is reduced below the 
non-accreting case, but only by about 25\%. 
Our main result is summarized in Fig.~\ref{fig:dota-mdot}, which shows that the planet migration does not depend on the viscous inflow speed
of the disk material. For small disk masses ($M_\mathrm{D}/M_\mathrm{J} < 0.2$) only, a Jupiter mass planet moves slower than the disk material. 
For higher disk masses (given by higher $\dot{m}$ in our calculations), values and subsequently larger surface densities, $\Sigma$, the inward migration becomes faster as well.

An important result of our simulations is the finding that during the migration of the planet the normalized torques
remain constant, a feature seen by \citet{duffell2014type}. 
In principle, it might be possible to extract approximate analytical relationships for the slow down factors $f_\mathrm{mig}$ 
and $f_\mathrm{still}$, which would then be useful to calculate theoretical migration tracks to be used, for example, in population synthesis models.
To do this, more elaborate parameter studies will be necessary in the future.  

Very qualitatively the shape of the curve is similar to that found by \citet{duffell2014type} by a complementary method, but we do not see signs of saturation.
Our migration rate is typically faster than theirs up to a factor of 2, but note that their results have been obtained for a spatially constant 
viscosity, which implies a spatially variable $\alpha$-value. Because of their special technique of pulling the planet through the disk and
measuring the torque after reaching equilibrium, it is not know at which distance of the planet this occurs. Hence, the value of $\alpha$
is not known and the curves cannot directly be compared to each other as the torque depends on the value of $\alpha$.
\citet{edgar2007giant,edgar2008type} also used constant viscosity. He also finds that the migration rate depends on the disk mass and the migration timescales are comparable to those we found.

Our findings are also in agreement with those of \citet{2000MNRAS.318...18N} who analysed the migration of a planet in a global disk and constant kinematic viscosity. 
They inferred a timescale for migration of a massive planet from its starting location at 5\,au to 2.5\,au of about 
2500 orbits, but did not compare this to the viscous accretion rate.

For sufficiently 
low planetary mass, $M_\mathrm{P} \lesssim M_\mathrm{D}$, but with $M_\mathrm{P}$ still large enough to open a partial gap,
the planets can enter the very fast type III migration regime where the migration timescale
becomes very short, of the order of about 100 dynamical times.
The conditions necessary for type III migration \citep{2003ApJ...588..494M} indicate that is may be relevant in the early phases of planet formation with larger disk
masses, or in more massive self-gravitating disks.

For low disk masses $M_\mathrm{D}/M_\mathrm{J} < 0.2$, 
the migration rate becomes lower than type II migration. These small disk masses occur
only during the end phases of the planet formation process when the accretion rate has already reduced significantly. 

In this work, we studied only 2D disks and found that disk material can always cross the gap region as required by the migration speed of the
planet. For massive, Jupiter type planets the averaged gap profile is identical for 2D and 3D disks \citep{2001ApJ...547..457K} and
one may expect very similar mass flow rates across the gap region, and hence similar migration accretion rates.
The additional assumption of locally isothermal disks
is not very restrictive as well because for massive planets that do open significant gaps, the dynamical behaviour is very similar to the isothermal
case as has been shown by \citet{2010A&A...523A..30B} for full 3D simulations of radiative and isothermal disks.

Nevertheless, it may be interesting to perform in the future 3D simulations of embedded, massive planets in radiative disks and analyse their migration
properties. Those simulations are also important to understand better the mass growth of planets because the
inclusion of the 3rd dimension may affect the mass accretion onto the planet, while it only mildly affects the migration.

Our results have consequence for population synthesis models in that a modification of the assumed type II migration speed is required for
the models, based on the torques acting on the planets. Possibly, more elaborate models that cover a larger parameter space will
allow the construction of suitable fit formulae for the migration of massive planets in the future, but this is beyond the scope of the present work.

Another point to consider is the angular momentum balance of the planet. We have taken out mass from the planet's Roche lobe, but
have not added this to the planet mass. In addition to the mass, the angular momentum of this material, which is gained
by the planet, has to be considered. A fraction of this will change the orbital angular momentum, which will influence the migration of the planet.
Future simulations need to consider this effect as well.

\begin{acknowledgements}
We thank Bertram Bitsch for very fruitful discussions and acknowledge the constructive and helpful comments from the referee to improve this paper.
The calculations were performed on systems of bwGRiD, the computer grid of the Baden W\"urttemberg state.
The work of Christoph D\"urmann is sponsored with a scholarship of the Cusanuswerk.

\end{acknowledgements}

\bibliographystyle{aa} 
\bibliography{migration,kley} 

\end{document}